\documentclass[aps,prd,nofootinbib,preprint,preprintnumbers,showpacs]{revtex4}
\usepackage[dvipdfmx]{graphicx}
\usepackage{amsmath,amssymb,color}
\usepackage{ulem}
\newcommand{\VEV}[1]{\langle #1 \rangle}
\newcommand{\tr}{\mathrm{tr}}
\newcommand{\Tr}{\mathrm{Tr}}
\newcommand{\diag}{\mathrm{diag}}
\begin{document}
\title{Revisiting Vector-like Quark Model with Enhanced Top Yukawa Coupling}
\author{Michio Hashimoto}
%\email{michioh@isc.chubu.ac.jp}
\affiliation{Chubu University, 1200 Matsumoto-cho, Kasugai-shi,  Aichi, 487-8501, Japan}
\date{\today}
\preprint{}
\begin{abstract}
We revisit a scenario with an enhanced top yukawa coupling
in vector-like quark (VLQ) models,
where the top yukawa coupling is larger than the standard model value
and the lightest VLQ has a negative yukawa coupling.
We find that the parameter space satisfying the LHC bounds of 
the Higgs signal strengths consistently with the precision measurements 
is rather wide.
Because the Lagrangian parameters of the yukawa couplings are large,
such scenario can be realized in some strongly interacting theories.
It also turns out that there is a noticeable relation between 
the contributions of the triangle and box diagrams in the $gg \to hh$ process
by using the lowest order of the $1/M$ expansion where $M$ is 
the heavy mass running in the loops.
\end{abstract}
\maketitle

\section{Introduction}

The LHC experiments have discovered a Higgs boson and revealed that 
its properties are similar to that of 
the Standard Model (SM)~\cite{Khachatryan:2016vau}.
It is thus essential to explore any signatures of physics beyond the SM (BSM).
One of the hint is that the observed signal strength of 
$pp \to t\bar{t}h$ channel is deviated from the SM value about twice,
$\mu_{tth}=2.3\raisebox{-0.5ex}{$\stackrel{+0.7}{\mbox{\scriptsize $-0.6$}}$}$
for the Run 1 combined data~\cite{Khachatryan:2016vau}, 
$\mu_{tth}=1.8\raisebox{-0.5ex}{$\stackrel{+0.7}{\mbox{\scriptsize $-0.7$}}$}$
for the ATLAS Run 2~\cite{ATLAS:2016axz}, and
$\mu_{tth}=1.5\raisebox{-0.5ex}{$\stackrel{+0.5}{\mbox{\scriptsize $-0.5$}}$}$
for the CMS Run 2 in the multilepton final states\footnote{
A combined result of the CMS Run 2 has not yet been reported.
For other decay channels,
$\mu_{tth}=1.91\raisebox{-0.5ex}
 {$\stackrel{\tiny +1.5}{\mbox{\fontsize{7.0pt}{0pt}\selectfont $-1.2$}}$}$
%for the CMS Run 2 
in the $h \to \gamma\gamma$ decay channel~\cite{CMS:2016ixj}, 
$\mu_{tth}=-0.19\raisebox{-0.5ex}
 {$\stackrel{\tiny +0.80}{\mbox{\fontsize{7.0pt}{0pt}\selectfont $-0.81$}}$}$
%for the CMS Run 2 
in the $h \to b\bar{b}$ decay channel~\cite{CMS:2016zbb}, and
$\mu_{tth}=0.00\raisebox{-0.5ex}
 {$\stackrel{\tiny +1.19}{\mbox{\fontsize{7.0pt}{0pt}\selectfont $-0.00$}}$}$
%for the CMS Run 2 
in the $h \to ZZ \to 4 \ell$ channel~\cite{CMS:2017jkd}. 
}~\cite{CMS:2017vru}, 
%in the process of $pp \to t\bar{t}H \to 
although the uncertainties are still large. 

These experiments provide a reason for considering models based on 
strongly interacting theories.
In this direction\footnote{
Although the top condensate model~\cite{Miransky:1988xi}
and the chiral fourth generation~\cite{4family} directly predict
large yukawa couplings, they had been severely constrained.
}, 
widely studied are vector-like quark (VLQ) models~\cite{Lavoura:1992np,Anastasiou:2009rv,Aguilar-Saavedra:2013qpa,Cacciapaglia:2011fx,Cacciapaglia:2010vn,Alok:2015iha,Alok:2014yua,Cacciapaglia:2015ixa,Angelescu:2015kga,Biekotter:2016kgi,Chen:2017hak}, 
the minimal composite Higgs models (MCHMs)~\cite{Agashe:2004rs,Contino:2006qr},
and the Little Higgs models~\cite{ArkaniHamed:2002qy,Schmaltz:2005ky}.
We easily find, however, 
the top yukawa coupling is always suppressed in 
the VLQ model having only one up-type quark~\cite{Aguilar-Saavedra:2013qpa}. 
For example, introducing the VLQ $U_{L,R}$ having $+2/3$ electric charge 
as in the top-seesaw model~\cite{TSS}, 
the top yukawa coupling is modified as $c_L^2 g_{\bar{t}t h}^{\rm SM}$,
where $c_L \equiv \cos \theta_L$ represents the cosine of the mixing angle
between $t_L$ and $U_L$, and we defined the SM top yukawa coupling by
$g_{\bar{t}t h}^{\rm SM} = m_t/v$ with $m_t$ and $v$ being 
the top mass and the vacuum expectation value (VEV) of the Higgs field,
respectively.
It is the case\footnote{
Quite recently, it is shown that the MCHMs with the fermions of
the ${\bf 5}+{\bf 10}$ or ${\bf 14}$ representations
can have the enhanced or suppressed $tth$ coupling~\cite{Liu:2017dsz}.
}
for the MCHMs such as MCHM4, MCHM5, and MCHM10 
where the fermions are embedded in the spinorial,
${\bf 5}$, and ${\bf 10}$ representations of $SO(5)$, 
respectively~\cite{Espinosa:2010vn,Carena:2014ria}. 
Nevertheless one should not jump to a conclusion:
A simple model is effective for a benchmark, but 
it might be misguided if simplified too much.

In this paper, we reconsider a scenario that 
the top yukawa coupling is larger than the SM value
by ${\cal O}(10\%)$, 
and the lightest VLQ with the mass around 1~TeV has a negative yukawa coupling 
of the order of $-m_t/v$, 
introducing more than one up-type 
VLQ~\cite{Cheng:2014dwa,Angelescu:2015kga,Cacciapaglia:2015ixa}.
In our scenario, owing to the cancellation among the yukawa couplings, 
the Higgs signal strengths can be consistent with the experiments.
A similar analysis\footnote{
In the framework of the two Higgs doublet model, 
the cancellation mechanism via the light stop was considered 
in Ref.~\cite{Badziak:2016exn,Badziak:2016tzl}. 
See also Ref.~\cite{Das:2017scg}. 
}
was performed in Ref.~\cite{Angelescu:2015kga}. 
Although the allowed region looked narrow in Ref.~\cite{Angelescu:2015kga}, 
we find that our scenario is possible in a rather wide parameter space. 

We numerically show that our scenario is realized, roughly speaking,
when the Lagrangian parameters $y_{ij}$ of the yukawa interactions 
are large, say, $|y_{ij}| \gtrsim 2$.
This may suggest the existence of the underlying strongly interacting 
models where the dynamically generated yukawa couplings are typically
around $3 \sim 5$~\cite{TSS}.
As for the di-Higgs production process $gg \to hh$~\cite{Kniehl:1995tn,Falkowski:2007hz,Low:2009di,Grober:2010yv,Gillioz:2012se,Dawson:2012mk,Chen:2014xwa,Grober:2016wmf},
we find a noticeable relation between the contributions of 
the triangle and the box diagrams in the lowest order of the $1/M$ expansion,
where $M$ is the relevant heavy mass running 
in the loops.
The di-Higgs production process may give 
information on the off-diagonal yukawa couplings.

The paper is organized as follows:
In Sec.~\ref{Sec-model}, we introduce the VLQ model. 
In Sec.~\ref{Sec-analysis}, we first describe the existence proof
of our scenario in an analytical approach, and next show a numerical 
calculation.
Sec.~\ref{Sec-summary} is devoted to summary.
In Appendix A, the oblique parameters~\cite{Peskin:1990zt} 
in our model are presented.
Analytical expressions of the triangle and the box contributions to 
the $gg \to hh$ process in the lowest order of the $1/M$ expansion 
are given in Appendix B.

\section{Vector-like quark model}
\label{Sec-model}

\begin{table}
 \begin{center}
$$
  \begin{array}{c|ccc} \hline
    & SU(3)_c & SU(2)_W & U(1)_Y \\[1mm] \hline
   q_L=(t,b)_L & {\bf 3} & {\bf 2} & \frac{1}{6} \\[2mm]
   t_R & {\bf 3} & {\bf 1} & \frac{2}{3} \\[2mm]
   b_R & {\bf 3} & {\bf 1} & -\frac{1}{3} \\[1mm] \hline
   Q_{L,R}=(X,T)_{L,R} & {\bf 3} & {\bf 2} & \frac{7}{6} \\[2mm]
   U_{L,R} & {\bf 3} & {\bf 1} & \frac{2}{3} \\[1mm] \hline
  \end{array}
$$
  \end{center}
  \caption{Charge assignment of the VLQ model.
   \label{charge}}
\end{table}

Let us introduce two types of the VLQ's,
$U_{L,R}$ and $Q_{L,R}=(X,T)_{L,R}$, 
having the hypercharges $\frac{2}{3}$ and $\frac{7}{6}$, respectively.
(See also Table~\ref{charge}.)
Because of no mixing between the bottom quark and VLQ's,
the flavor constraints such as $Z \to b\bar{b}$, etc. can be suppressed 
in this model.
Assuming one Higgs doublet model, the mass terms and the yukawa interactions
are
\begin{eqnarray}
  {\cal L}^{}_Y &=&
%  - y^{}_{11} \bar{q}_L \tilde{H} t_R
  - y^{}_{13}\bar{q}_L \tilde{H} U_R
  - y^{}_{21} \bar{Q}_L H t_R - y^{}_{23} \bar{Q}_L H U_R
  - y^{}_{32} \bar{U}_L H^\dagger Q_R + \mbox{(h.c)}, \\
  {\cal L}^{}_{\rm VM} &=&
  - m^{}_{22} \bar{Q}_L Q_R - m^{}_{33} \bar{U}_L U_R
  - m^{}_{31} \bar{U}_Lt_R + \mbox{(h.c)}, 
\end{eqnarray}
with $q_L=(t,b)_L$, 
%$Q_{L,R}=(X,T)_{L,R}$, 
and $\tilde{H} \equiv i\tau_2 H^*$.
The SM term of $y_{11}\bar{q}_L \tilde{H} t_R$ was rotated away via
the $t_R$--$U_R$ mixing like in the top seesaw model~\cite{TSS},
while $m_{31}$ is removed 
in literature~\cite{Angelescu:2015kga,Cacciapaglia:2015ixa}. 
We here abbreviated the SM part such as the light quark sector,
gauge kinetic terms, etc..

After the electroweak symmetry breaking (EWSB), 
the mass matrix is then
\begin{equation}
  {\cal L}_M = - (\bar{t}_L\; \bar{T}_L\;\bar{U}_L) \; {\cal M} \;
  \left(
  \begin{array}{c}
    t_R \\ T_R \\ U_R
  \end{array}
  \right) - m^{}_{22} \bar{X}_L X_R  
\end{equation}
with $\VEV{H}=(0,\;\frac{v}{\sqrt{2}})^T$, $v=246$~GeV, and
\begin{equation}
  {\cal M} \equiv 
  \frac{v}{\sqrt{2}} {\mathbf Y} \oplus  {\mathbf M}
  = \frac{v}{\sqrt{2}} 
   \left(
    \begin{array}{ccc}
     0  & 0 & y^{}_{13} \\ y^{}_{21} & 0 & y^{}_{23} \\ 0 & y^{}_{32} & 0
    \end{array}
   \right) +
   \left(
    \begin{array}{ccc}
      0 & 0 & 0 \\ 0 & m^{}_{22} & 0 \\ m^{}_{31} & 0 & m^{}_{33}
    \end{array}
   \right) ,
   \label{mass-matrix-general}
\end{equation}
where the mass of the $X$ quark with $+5/3$ electric charge is not
affected by the EWSB, i.e., $M_X = m_{22}$. 
We diagonalize ${\cal M}$ by
\begin{equation}
  {\cal M}_{\rm diag} = V_L^\dagger {\cal M} V_R, \quad
  {\cal M}_{\rm diag} = \diag (m_1,m_2,m_3),
\end{equation}
with $0 \leq m_1 \leq m_2 \leq m_3$, and
\begin{equation}
  \left(\begin{array}{@{}c@{}}
    t_{L,R} \\ T_{L,R} \\ U_{L,R}
  \end{array}\right) = V_{L,R}
  \left(\begin{array}{@{}c@{}}
    t'_{L,R} \\ T'_{L,R} \\ U'_{L,R}
  \end{array}\right) ,
\end{equation}
where $(t,T,U)_{L,R}$ and $(t',T',U')_{L,R}$ represent
the gauge and mass eigenstates, respectively.
Each up-type quark mass is identified by $m_t=m_1$, $M_T=m_2$, and $M_U=m_3$.
The yukawa coupling matrix ${\mathbf G}^h$ in the mass eigenstates is given by
\begin{eqnarray}
  {\cal L}_Y 
  &=& - h \; (\bar{t}'_L\; \bar{T}'_L\;\bar{U}'_L) \; {\mathbf G}^h \;
  \left(
  \begin{array}{c}
    t'_R \\ T'_R \\ U'_R
  \end{array}
  \right),
\end{eqnarray}
with
\begin{equation}
   {\mathbf G}^h = \frac{1}{\sqrt{2}}\, V_L^\dagger {\mathbf Y} V_R \, .
\end{equation}

\section{Analytical and numerical studies}
\label{Sec-analysis}

\subsection{Analytical study with crude approximation}
\label{crude-app}

We schematically show our scenario having ${\mathbf G}^h_{11} > m_t/v$
and ${\mathbf G}^h_{22} < 0$ is possible in an analytical approach.
For this purpose, we employ a crude approximation in this subsection.
A numerical study without such approximation will be shown 
in the next subsection.

Let us take the mass matrix as a symmetric one,
\begin{equation}
  {\cal M} = M_X 
  \left(
   \begin{array}{ccc}
      0 & 0 & \epsilon \\ 0 & 1 & \xi \\ \epsilon & \xi & a
    \end{array}
   \right),
  \label{mass-matrix}
\end{equation}
where we scaled the mass matrix by $M_X=m_{22}$. 
%From the viewpoint of calculablity, 
For the perturbation theory, the parameters arising from the yukawa couplings
should not be so large, i.e., $\epsilon^2, \xi^2 \ll 1$. 
We also assume $a \geq 1$.
%$a \gtrsim 1$.
We diagonalize ${\cal M}$ by the matrices of
\begin{equation}
  V_L = (\vec v_t \; \vec v_T \; \vec v_U ), \quad
  V_R = (-\vec v_t \; \vec v_T \; \vec v_U )=V_L \diag(-1,1,1), 
\end{equation}
where the first, the second and the third components 
($v_{t1}^{}$, $v_{T2}^{}$ and $v_{U3}^{}$) of $\vec v_t$, $\vec v_T$ 
and $\vec v_U$ are order of unity, 
\begin{equation}
  v_{t1}^{} \equiv \vec v_t^{\:(1)} = {\cal O}(1), \quad
  v_{T2}^{} \equiv \vec v_T^{\:(2)} = {\cal O}(1), \quad
  v_{U3}^{} \equiv \vec v_U^{\:(3)} = {\cal O}(1), 
\end{equation}
respectively, and then obtain the mass eigenvalues,
\begin{equation}
  \frac{m_t}{M_X} \equiv \lambda_1 = {\cal O}\left(\frac{\epsilon^2}{a}\right),
  \quad
  \frac{M_T}{M_X} \equiv \lambda_2 < 1, \quad
  \frac{M_U}{M_X} \equiv \lambda_3 > a \, . 
\end{equation}
In general, by taking the trace and the determinant, we find
\begin{equation}
  -\lambda_1 + \lambda_2 + \lambda_3 = 1+a, \quad
   \lambda_1  \lambda_2  \lambda_3 = \epsilon^2, \quad
   -G_{tt}^h + G_{TT}^h + G_{UU}^h = 0 ,
  \label{formula-gen1}
\end{equation}
where we defined the diagonal components of the yukawa couplings 
in the mass basis as ${\mathbf G}^h_{11,22,33} \equiv G_{tt,TT,UU}^h$.
More explicitly, the yukawa couplings of $G_{tt}^h$ and $G_{TT}^h$
are given by
\begin{eqnarray}
  G_{tt}^h &=& \frac{m_t}{v}\bigg[\, 1 + \frac{v_{t1}^2}{(1+\lambda_1)\epsilon^2}
  \Big\{\,(a\lambda_1 - \epsilon^2)(2+\lambda_1) + \lambda_1^2\,\Big\}\,\bigg],
  \label{formula-g11} \\
  G_{TT}^h &=& - (1-\lambda_2) \frac{M_X}{v}\bigg[\, 
  1 + \frac{(1-\lambda_2) (a-\lambda_2)}{\xi^2}\,\bigg] v_{T2}^2 < 0,
  \label{formula-g22}
\end{eqnarray}
and the situation of $G_{tt}^h > m_t/v$ is realized when
\begin{equation}
  \xi^2 > \frac{1+\lambda_0}{2}\Big(\epsilon^2 - (a-1)\lambda_0\Big),
\end{equation}
with
\begin{equation}
  \lambda_0 \equiv \frac{4\epsilon^2}{2a-\epsilon^2 +
    \sqrt{(2a-\epsilon^2)^2+8\epsilon^2 (1+a)}} \, .
\end{equation}

An analytic solution is frequently useful.
Let us take $\xi = \frac{\epsilon}{a} \sqrt{a+\epsilon^2}$, for example.
In this case, we find that the eigenvalues are
\begin{equation}
  \lambda_1 = \frac{\epsilon^2}{a}, \quad
  \lambda_2 = 1-\delta^2, \quad
  \lambda_3 = a + \frac{\epsilon^2}{a} + \delta^2,
\end{equation}
with
\begin{equation}
  \delta^2 \equiv
  \frac{2\epsilon^2}
       {a(a-1)+\epsilon^2+\sqrt{\{a(a-1)+\epsilon^2\}^2+4a\epsilon^2}} ,
\end{equation}
and the corresponding eigenvectors are
\begin{equation}
  \vec{v}_{t} = v_{t1}^{} \left(
   \begin{array}{c}
      1 \\ \frac{\epsilon^2}{a\sqrt{a+\epsilon^2}} \\ 
       -\frac{\epsilon}{a}
   \end{array}
   \right), \quad
  \vec{v}_{T} = v_{T2}^{} \left(
   \begin{array}{c}
      - \frac{\lambda_3 \delta^2}{\sqrt{a+\epsilon^2}} \\ 1 \\
      - \frac{\epsilon}{(\lambda_3-1)\sqrt{a+\epsilon^2}}
   \end{array}
   \right), \quad
  \vec{v}_{U} = v_{U3}^{} \left(
   \begin{array}{c}
      \frac{\epsilon (1-\delta^2)}{a} \\
      \frac{\epsilon\sqrt{a+\epsilon^2}}{a(\lambda_3-1)} \\ 1
   \end{array}
   \right) ,
\end{equation}
with
\begin{eqnarray}
  v_{t1}^{-1} &=&
  \sqrt{1 + \frac{\epsilon^2(a+2\epsilon^2)}{a^2(a+\epsilon^2)}}, \\
  v_{T2}^{-1} &=& 
  \sqrt{1 + \frac{\delta^4(a^2+\epsilon^2\lambda_3^2)}
                 {\epsilon^2(a+\epsilon^2)}}, \\
  v_{U3}^{-1} &=& 
  \sqrt{1 + \frac{\epsilon^2}{a^2}
        \left((1-\delta^2)^2 + \frac{a+\epsilon^2}{(\lambda_3-1)^2}\right)}\,.
\end{eqnarray}
The following relations might be useful:
\begin{equation}
  (1-\lambda_2)(\lambda_3-1) = \delta^2 (\lambda_3-1) = \frac{\epsilon^2}{a}, 
  \qquad
  \lambda_2 \lambda_3 = \lambda_3 (1-\delta^2) = a \, .
\end{equation}
Substituting the above results for 
Eqs.~({\ref{formula-gen1})--(\ref{formula-g22}), we explicitly obtain
\begin{eqnarray}
  G_{tt}^h &=& \frac{m_t}{v}\bigg[\, 1 + 
  \frac{a \epsilon^2}{a^3 + a(a+1)\epsilon^2+2\epsilon^4}\,\bigg]>\frac{m_t}{v},
  \label{formula-anal-g11} \\
  G_{TT}^h &=& - \frac{M_X}{v}
  \delta^2
  \left(2-\frac{a\delta^2 + \epsilon^2}{a+\epsilon^2}\right)v_{T2}^2 < 0, \\
  % \frac{\epsilon^2 (2a-a\delta^2+\epsilon^2)}
  %      {(a+\epsilon^2)\{a (a-1)+\epsilon^2\}
  %       + \delta^2\{2a^2+\epsilon^2 (\lambda_3^2+a)\}} < 0, \\
%
 G_{UU}^h &=& G_{tt}^h - G_{TT}^h \, . 
\end{eqnarray}

In this way, our scenario can be realized in the framework of 
the VLQ model.
The parameter space should be constrained by the $S$ and $T$-parameters,
however.

\subsection{Numerical study without approximation}

Without assuming the symmetric mass matrix (\ref{mass-matrix}),
we now calculate numerically the signal strengths in our model:
\begin{eqnarray}
  \mu_F^{VV} \equiv \mu_{\rm ggF + ttH}^{VV}
  &=& \frac{(\sigma_{ggF} + \sigma_{ttH}) \mbox{Br}^{VV}}
         {(\sigma_{ggF} + \sigma_{ttH})_{\rm SM}\mbox{Br}_{\rm SM}^{VV}}
% &\simeq& \frac{(15.0 + 19.2)\kappa_g^2 + (0.086+0.129) \kappa_t^2}
%               {(15.0 + 19.2)+(0.086+0.129)}, \\
 \simeq \kappa_g^2, \\
  \mu_F^{\gamma\gamma} \equiv \mu_{\rm ggF + ttH}^{\gamma\gamma}
  &=& \frac{(\sigma_{ggF} + \sigma_{ttH}) \mbox{Br}^{\gamma\gamma}}
         {(\sigma_{ggF} + \sigma_{ttH})_{\rm SM} \mbox{Br}_{\rm SM}^{\gamma\gamma}}
  \simeq \kappa_g^2 \kappa_\gamma^2, 
%
%  \mu_V^{\gamma\gamma} \equiv \mu_{\rm VBF + VH}^{\gamma\gamma}
%  &=& \frac{(\sigma_{\rm VBF} + \sigma_{VH}) \mbox{Br}^{\gamma\gamma}}
%         {(\sigma_{\rm VBF} + \sigma_{VH})_{\rm SM}\mbox{Br}_{\rm SM}^{\gamma\gamma}}
% \simeq \kappa_\gamma^2, 
\end{eqnarray}
where $VV=WW$ and $ZZ$, 
and the scaling factors $\kappa_{g,\gamma,f}$ are defined by
\begin{eqnarray}
  \kappa_g &=&
   \frac{\kappa_t^{} A_{\frac{1}{2}} (x_t)
         + \frac{m_t}{M_T} \kappa_T^{} A_{\frac{1}{2}} (x_T)
         + \frac{m_t}{M_U} \kappa_U^{} A_{\frac{1}{2}} (x_U)}
        {A_{\frac{1}{2}} (x_t)}, \\
  \kappa_\gamma &=&
   \frac{A_1(x_W) + \frac{4}{3} (\kappa_t^{} A_{\frac{1}{2}} (x_t)
         + \frac{m_t}{M_T} \kappa_T^{} A_{\frac{1}{2}} (x_T)
         + \frac{m_t}{M_U} \kappa_U^{} A_{\frac{1}{2}} (x_U))}
        {A_1 (x_W) + \frac{4}{3} A_{\frac{1}{2}} (x_t)}, \\
  \kappa_t &=& G_{tt}^h/g_{tth}^{\rm SM}, \qquad
   g_{tth}^{\rm SM} \equiv \frac{m_t}{v},\\
  \kappa_T &=& G_{TT}^h/g_{tth}^{\rm SM}, \qquad
  \kappa_U = G_{UU}^h/g_{tth}^{\rm SM}, 
\end{eqnarray}
with $x_i \equiv m_h^2/(4m_i^2)$.
The loop functions for spin 1 and $1/2$ are represented by
$A_1(x)$ and $A_{1/2}(x)$, respectively~\cite{Gunion:1989we,Djouadi:2005gi},
\begin{eqnarray}
  A_1 (x) &=& - \frac{1}{x^2}\bigg[\,2x^2 + 3 x + 3 (2x-1) f(x)\,\bigg], \\
  A_{\frac{1}{2}} (x) &=& \frac{2}{x^2}\bigg[\,x + (x-1) f(x)\,\bigg], 
%  A_0 (x) &=& - \frac{1}{x^2}\bigg[\,x - f(x)\,\bigg],
\end{eqnarray}
with
\begin{equation}
  f(x) \equiv \left\{
    \begin{array}{lcc}
      \arcsin^2\sqrt{x} & \mbox{for} & x \leq 1, \\[2mm]
      {\displaystyle 
       -\frac{1}{4} \bigg[\,
         \ln\frac{1+\sqrt{1-x^{-1}}}{1-\sqrt{1-x^{-1}}} - i\pi\,\bigg]^2
      } & \mbox{for} & x > 1 \, . \\
    \end{array}
    \right. 
\end{equation}
% Because of $\kappa_V = \kappa_b = \kappa_\tau = \kappa_c = \kappa_\mu=1$
% in our model, the signal strength $\mu_V^{WW/ZZ}$ is SM-like,
% \begin{equation}
%   \mu_V^{WW/ZZ} \equiv \mu_{\rm VBF + VH}^{WW/ZZ}
%   = \frac{(\sigma_{\rm VBF} + \sigma_{VH}) \mbox{Br}^{WW/ZZ}}
%          {(\sigma_{\rm VBF} + \sigma_{VH})_{\rm SM}\mbox{Br}_{\rm SM}^{WW/ZZ}}
%   \simeq 1 \, .
% \end{equation}
In our model, the scaling factor $\kappa_V$ of the $hWW$ and $hZZ$ couplings
is SM-like, $\kappa_V=1$. 
Since we do not change the down quark and lepton sectors, 
the scaling factors of the bottom and tau are also
$\kappa_b=\kappa_\tau=1$.

By using the results of the LHC Run~1 via the six-parameter fit shown 
in Ref.~\cite{Khachatryan:2016vau},
\begin{eqnarray}
  \mu_V/\mu_F &=&
   1.09 \raisebox{-0.5ex}{$\stackrel{+0.36}{\mbox{\scriptsize $-0.28$}}$}, \\ 
  \mu_F^{\gamma\gamma} &=& 
   1.10 \raisebox{-0.5ex}{$\stackrel{+0.23}{\mbox{\scriptsize $-0.21$}}$}, \\ 
  \mu_F^{ZZ} &=& 
   1.27 \raisebox{-0.5ex}{$\stackrel{+0.28}{\mbox{\scriptsize $-0.24$}}$}, \\ 
  \mu_F^{WW} &=& 
   1.06 \raisebox{-0.5ex}{$\stackrel{+0.21}{\mbox{\scriptsize $-0.18$}}$}, \\ 
  \mu_F^{\tau\tau} &=& 
   1.05 \raisebox{-0.5ex}{$\stackrel{+0.33}{\mbox{\scriptsize $-0.27$}}$}, \\ 
  \mu_F^{bb} &=& 
   0.64 \raisebox{-0.5ex}{$\stackrel{+0.37}{\mbox{\scriptsize $-0.28$}}$} ,
\end{eqnarray}
we read the $2\sigma$ constraints as
\begin{equation}
  0.79 <  \mu_F^{VV} < 1.48, \qquad 0.68 < \mu_F^{\gamma\gamma} < 1.56,
  \label{Run1-2sigma}
\end{equation}
because of $\mu_F^{WW}=\mu_F^{ZZ} \equiv \mu_F^{VV}$ in our model.
On the other hand, the best fit values of 
$(\sigma \cdot B)^{ZZ}_{ggF}$, $\sigma_{VBF}/\sigma_{ggF}$
and $B^{\gamma\gamma}/B^{ZZ}$ yield
$\mu_{ggF}^{ZZ} = 1.42 \raisebox{-0.5ex}{$\stackrel{+0.35}{\mbox{\scriptsize $-0.31$}}$}$
and 
$\mu_{ggF}^{\gamma\gamma}=0.67\raisebox{-0.5ex}{$\stackrel{+0.25}{\mbox{\scriptsize $-0.21$}}$}$
in the ATLAS Run~2~\cite{ATLAS:2016hru}.
The signal strengths in the CMS Run~2 are
$\mu_{ggF}^{ZZ} = 1.20 \raisebox{-0.5ex}{$\stackrel{+0.22}{\mbox{\scriptsize $-0.21$}}$}$~\cite{CMS:2017jkd} 
and 
$\mu_{ggF}^{\gamma\gamma}=0.77\raisebox{-0.5ex}{$\stackrel{+0.25}{\mbox{\scriptsize $-0.23$}}$}$~\cite{CMS:2016ixj}.
One should keep in mind that both of the Run~2 results 
for $\mu_{ggF}^{\gamma\gamma}$ are much smaller than that of the LHC Run~1.

The parameter space of the mass matrix (\ref{mass-matrix-general})
is constrained by the precision measurements~\cite{Peskin:1990zt}.
Especially, owing to the mixing among $t$, $T$ and $U$,
the $T$-parameter is potentially large.
We explicitly show the expression of the $S$ and $T$-parameters in our model 
in Appendix A~\cite{Lavoura:1992np,Anastasiou:2009rv}.
Fixing $U=0$, we impose the constrains~\cite{Olive:2016xmw}, 
\begin{equation}
  \Delta S = 0.07 \pm 0.08, \qquad \Delta T = 0.10 \pm 0.07 ,
  \label{ST-const}
\end{equation}
where $m_t=173.3$~GeV, $m_b=4.2$~GeV, and $m_h=125.1$~GeV~\cite{Olive:2016xmw}.

We now describe the numerical results.
In the following analysis, we take the Higgs mass,
the pole mass of the top, the $\overline{\rm MS}$ mass of the bottom, and 
the CKM matrix element for $t$ and $b$ as
$m_h=125.1$~GeV, $m_t^{\rm pole}=173.2$~GeV, 
$m_b^{\overline{\rm MS}}=4.2$~GeV, and $|V_{tb}|=0.95$, respectively.
The relation $m_1=m_t$ must hold.
Although strong couplings are acceptable in our scenario,
we may impose $|y_{ij}| < 5$ for the Lagrangian parameters in 
Eq.~(\ref{mass-matrix-general}).
Even in this case, there is still wide parameter space, 
as we will see below.
Considering the lower mass bound for the $T$ quark~\cite{ATLAS:2016btu}, 
we fix $M_T=1.2$~TeV and take the mass range for the heavier VLQ to
$1.5 \leq M_U \leq 3.5$ TeV.

The signal strengths of $\mu_F^{VV}$ and $\mu_F^{\gamma\gamma}$
are depicted in Figs.~\ref{muFW-fig} and \ref{muF-fig}.
For the red points, the $S,T$ constraints and 
the $2\sigma$ bounds (\ref{Run1-2sigma})
of the Higgs signal strengths are satisfied,
while the green points are outside of the $2\sigma$ bounds (\ref{Run1-2sigma}).
For the blue points in Figs.~\ref{muFW-fig} and \ref{muF-fig}, 
$G_{tt}^h > g_{tth}^{\rm SM}$ and $G_{TT}^h > 0$.
We did not plot the data with $G_{tt}^h < g_{tth}^{\rm SM}$ in our model,
although they exist.
We also show the results for MCHM4 and MCHM5,
where the scaling parameters are
$\kappa_V = \sqrt{1-\xi}$ for both and $\kappa_f=\sqrt{1-\xi}$ 
and $\kappa_f = (1-2\xi)/\sqrt{1-\xi}$ for MCHM4 and MCHM5, respectively,
with $\xi=v^2/f^2$ and $f$ being the typical scale of the MCHMs~\cite{Espinosa:2010vn,Carena:2014ria,Kanemura:2016tan,Liu:2017dsz,Sanz:2017tco}.
The $2\sigma$ constraint of the top yukawa coupling from 
the Run 1 combined data~\cite{Khachatryan:2016vau}, 
which reads $1.05 < \kappa_t < 1.92$,
is also shown in Figs.~\ref{muFW-fig} and \ref{muF-fig}
with the proviso that the Run 2 data do not restrict the top yukawa 
so much yet, within the $2\sigma$ bounds,
$0.63 < \kappa_t < 1.79$~(ATLAS Run2~\cite{ATLAS:2016axz})
and $0.71 < \kappa_t < 1.58$~(CMS Run2~\cite{CMS:2017vru}).
In passing, we comment that there is a parameter space inside of 
the $2\sigma$ bounds (\ref{Run1-2sigma}), even if we take $M_T=2.0$~TeV.
Although the window is closed at $M_T=2.4$~TeV under the condition
$|y_{ij}| < 5$,
the parameter space still exists even for $M_T=3.0$~TeV, 
if we allow $|y_{ij}| > 5$. 

In our scenario, we require $G_{tt}^h > g_{tth}^{\rm SM}$ and $G_{TT}^h < 0$
in order for the Higgs signal strengths to be consistent with 
the experiments.
For this cancellation mechanism among the diagonal yukawa couplings
in the gluon fusion process, we show the normalized yukawa couplings of 
$\kappa_{T,U} = G_{TT,UU}^h/g_{tth}^{\rm SM}$ vs 
$\kappa_t = G_{tt}^h/g_{tth}^{\rm SM}$ in Fig.~\ref{kT-kt}.
At the red points, the conditions of $G_{tt}^h > g_{tth}^{\rm SM}$ and 
$G_{TT}^h < 0$, the $2\sigma$ constraints (\ref{Run1-2sigma}),
and the $S,T$-constraints (\ref{ST-const}) are satisfied. 
On the other hand, the green points are outside of
the $2\sigma$ constraints (\ref{Run1-2sigma}).
We find that the cancellation mechanism works up to $\kappa_t \lesssim 1.4$. 

The Lagrangian parameters are important for the model-building.
We depict them in Fig.~\ref{fig-yij}.
The entry of $y_{21}$ can be either positive or negative.
The vanishing $y_{21}$ is also possible.
This is consistent with the analytical approach in the previous subsection.
Although $y_{23}$ barely takes negative or small positive values,
$y_{ij}$ except for $y_{21}$ are positive and large, say, $y_{ij} \gtrsim 2$,
in the wide parameter space. 
Thus the VLQ model discussed here might be provided from some underlying
theories based on strongly interacting systems.

In the end of this subsection,
we comment on the di-Higgs production via the gluon fusion.
The off-diagonal yukawa couplings can be extracted from 
the decay channels such as $T \to th$.
Also, they contribute to the box diagram of the di-Higgs production,
so that the $gg \to hh$ process may give us further information
on the model parameters.
In the lowest order of the $1/M$ expansion (LET),
%~\cite{Gillioz:2012se,Dawson:2012mk},
%the ratios to the SM results of the triangle and box diagrams are
the triangle and box contributions normalized by the SM values
are~\cite{Kniehl:1995tn,Falkowski:2007hz,Low:2009di,Gillioz:2012se,Dawson:2012mk,Chen:2014xwa}
%Grober:2010yv,
\begin{eqnarray}
  R_{gg \to h}^{\rm tri} &=& \frac{{\cal A}_{gg \to h}}{{\cal A}_{gg \to h}^{\rm SM}}
  = v \Tr ({\mathbf G}^h{\cal M}_{\rm diag}^{-1}), \label{def-Rtri} \\
  R_{gg \to hh}^{\rm box} &=&
  \frac{{\cal A}_{gg \to hh}^{\rm box}}{{\cal A}_{gg \to hh}^{\rm SM,box}}
  = v^2 \Tr ({\mathbf G}^h {\cal M}_{\rm diag}^{-1}
             {\mathbf G}^h {\cal M}_{\rm diag}^{-1}), 
  \label{def-Rbox}
\end{eqnarray}
respectively.
Note that $R_{gg \to h}^{\rm tri} \approx \kappa_g$, because of
$A_{\frac{1}{2}} (x_t) \simeq A_{\frac{1}{2}} (x_T) \simeq 
 A_{\frac{1}{2}} (x_U) \approx A_{\frac{1}{2}} (0) = 4/3$.
The numerical result is depicted in Fig.~\ref{Rtri-box}.
Also, we can analytically obtain the expressions of 
$R_{gg \to h}^{\rm tri}$ and $R_{gg \to hh}^{\rm box}$ 
in the general case (\ref{mass-matrix-general}) as
\begin{equation}
  R_{gg \to h}^{\rm tri} = 3 - 2 b_{22}, \quad
  R_{gg \to hh}^{\rm box} = 3-6b_{22} + 4b_{22}^2 =
  \Big(R_{gg \to h}^{\rm tri}\Big)^2 - 3 \Big(R_{gg \to h}^{\rm tri}-1\Big),
  \label{Rtri-Rbox-anal}
\end{equation}
where $b_{22}$ denotes the $(2,2)$ element of the dimensionless
mass matrix inverse $M_X {\cal M}^{-1}$.
See also Appendix \ref{app-b}.
This analytic result is shown as the blue curve in Fig.~\ref{Rtri-box}.
Under the symmetric mass matrix assumption (\ref{mass-matrix})
in the previous subsection,
we find $R_{gg \to h}^{\rm tri}=R_{gg \to hh}^{\rm box}=1$ owing to $b_{22}=1$.
It then turns out that the box contribution $R_{gg \to hh}^{\rm box}$ is 
decreasing with respect to the increasing triangle contribution
$R_{gg \to h}^{\rm tri}$ under the $2\sigma$ constraints (\ref{Run1-2sigma}).
Since the box is destructive in $gg \to hh$,
it means that the di-Higgs production is either much enhanced
or suppressed.
In Fig.~\ref{sigma_LET}, we show the ratio
$\sigma_{\rm LET}/\sigma_{\rm LET}^{\rm SM}$ of the total cross section of 
the Higgs pair production through $gg$ in $pp$ collisions normalized by 
the SM one under the LET approximation~\cite{Kniehl:1995tn,Falkowski:2007hz,Low:2009di,Gillioz:2012se,Dawson:2012mk,Chen:2014xwa}.
We took the LHC center of mass energy as $\sqrt{s}=14$~TeV and 
used the CT14 LO PDF set~\cite{Dulat:2015mca}. 
The renormalization scale ($\mu$) and the factorization scale ($Q$)
are chosen equal to the invariant mass ($M_{hh}=\sqrt{\hat{s}}$)
of the Higgs pair, $\mu=Q=M_{hh}$.
We find that the ratio can be increasing/decreasing about 40\%,
depending on the values of $R_{gg \to h}^{\rm tri} \approx \kappa_g$.
This might be striking, although one should take notice of inaccuracy
of the LET approximation.
A detailed analysis will be performed elsewhere.

\begin{figure}[t]
  \begin{center}
   \includegraphics[width=0.95\textwidth]{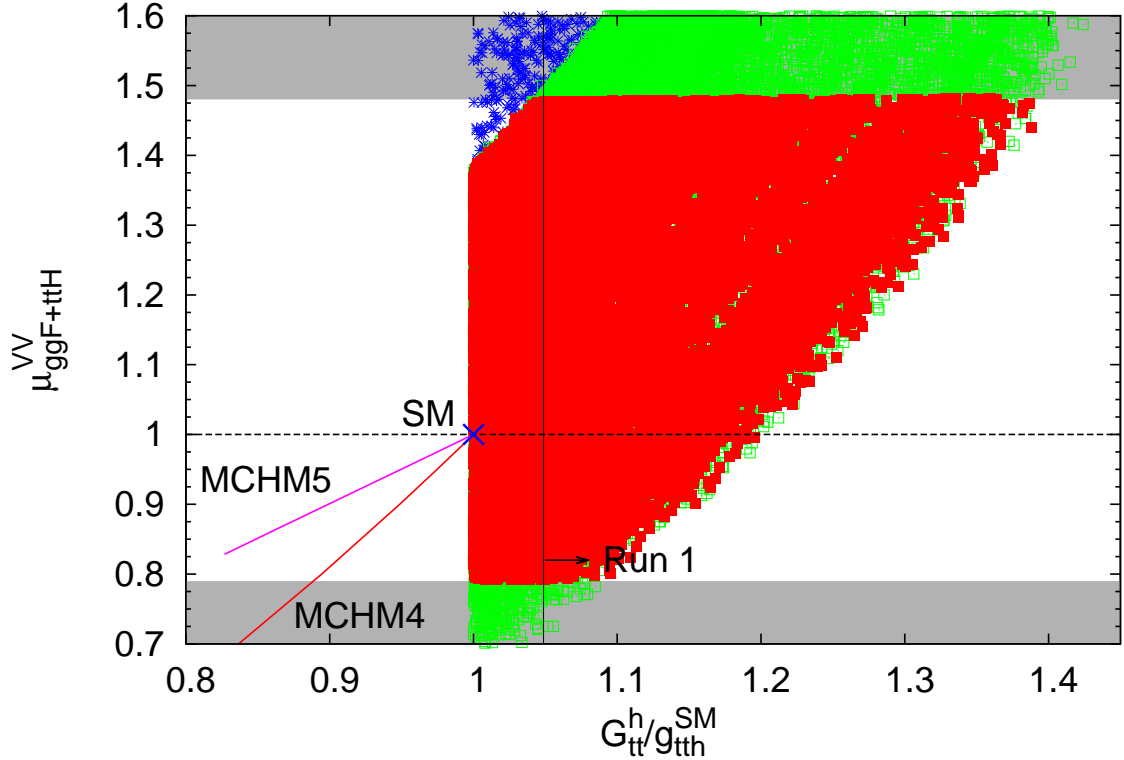}
  \end{center}
  \caption{$\mu_{ggF+ttH}^{VV}$ vs $G_{tt}^h/g_{tth}^{\rm SM}$.
   We fixed 
   %$V_{tb} = 0.95$, 
   $M_T=1.2$ TeV and took the mass range, $1.5 \leq M_U \leq 3.5$~TeV.
   The upper and lower shaded regions are outside of the $2\sigma$
   constraints (\ref{Run1-2sigma}).
   %of the Run 1 results~\cite{Khachatryan:2016vau}.
   The red points are inside of the $2\sigma$ constraints of the LHC Run 1.
   The green points satisfy only the conditions of $G_{tt}^h/g_{tth}^{\rm SM}>1$
   and $G_{TT}^h < 0$, and the $S,T$-constraints, while in the blue ones, 
   $G_{tt}^h/g_{tth}^{\rm SM}>1$ and $G_{TT}^h > 0$.
   We do not show the results with $G_{tt}^h/g_{tth}^{\rm SM} < 1$ in our model,
   although they exist.
   We also show the results for MCHM4 and MCHM5.
   \label{muFW-fig}}
\end{figure}

\begin{figure}[t]
  \begin{center}
   \includegraphics[width=0.95\textwidth]{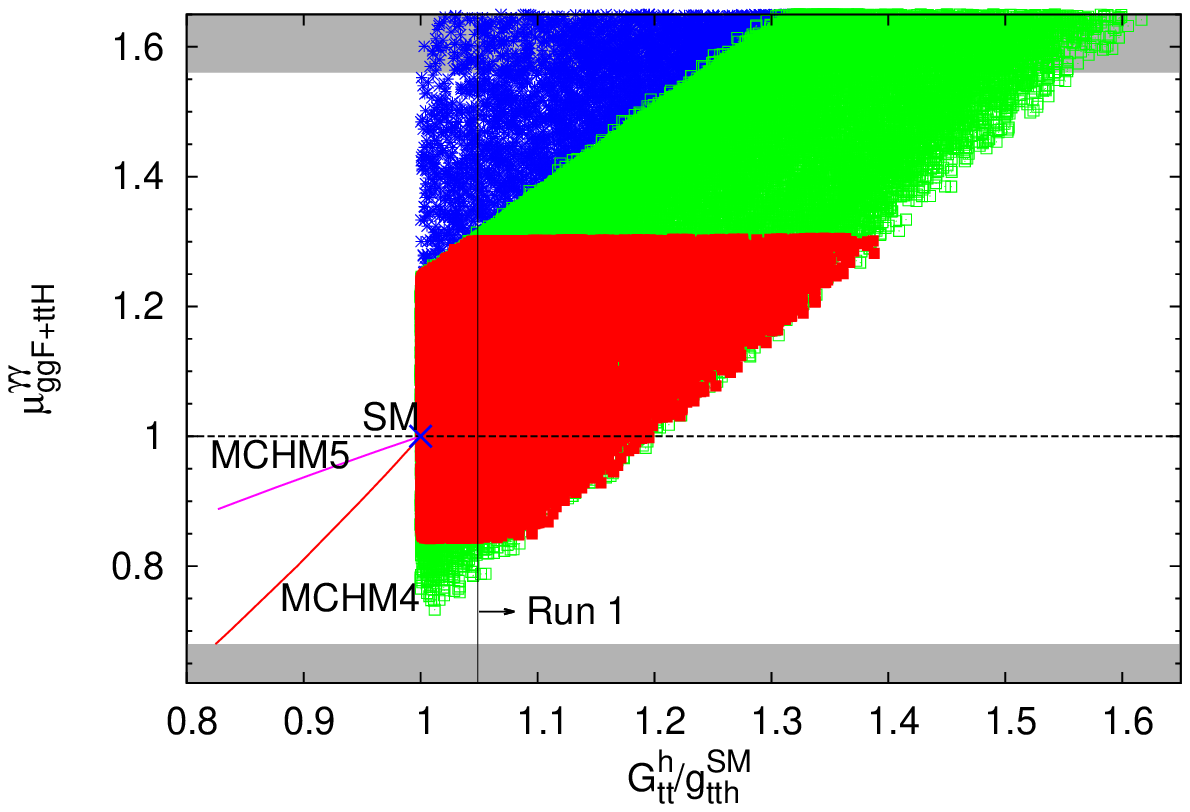}
  \end{center}
  \caption{$\mu_{ggF+ttH}^{\gamma\gamma}$ vs $G_{tt}^h/g_{tth}^{\rm SM}$.
   We fixed 
   %$V_{tb} = 0.95$, 
   $M_T=1.2$ TeV and took the mass range, $1.5 \leq M_U \leq 3.5$~TeV.
   The upper and lower shaded regions are outside of the $2\sigma$
   constraints (\ref{Run1-2sigma}).
   The red points are inside of the $2\sigma$ constraints of the LHC Run 1.
   The green points satisfy only the conditions of $G_{tt}^h/g_{tth}^{\rm SM}>1$
   and $G_{TT}^h < 0$, and the $S,T$-constraints, while in the blue ones, 
   $G_{tt}^h/g_{tth}^{\rm SM}>1$ and $G_{TT}^h > 0$.
   We do not show the results with $G_{tt}^h/g_{tth}^{\rm SM} < 1$ in our model,
   although they exist.
   We also show the results for MCHM4 and MCHM5.
   \label{muF-fig}}
\end{figure}

\begin{figure}[t]
  \begin{center}
   \includegraphics[width=0.4\textwidth]{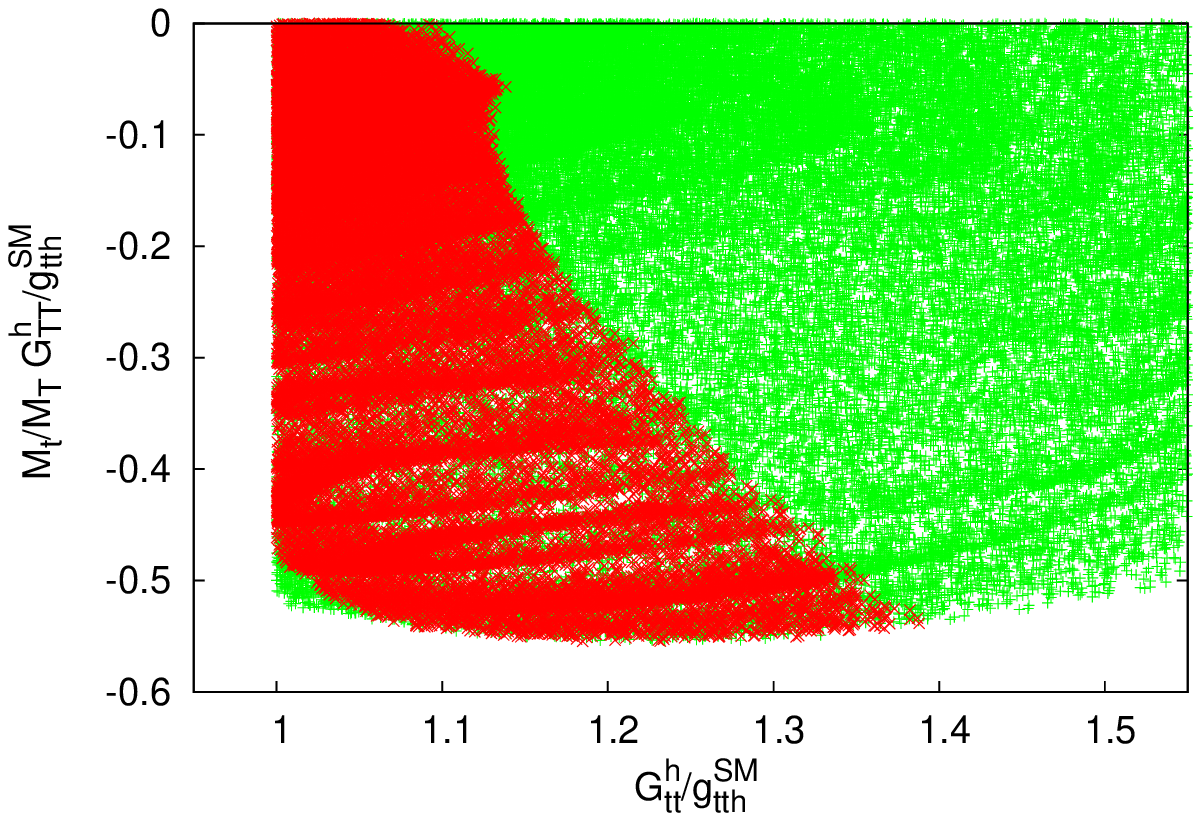}\qquad
   \includegraphics[width=0.4\textwidth]{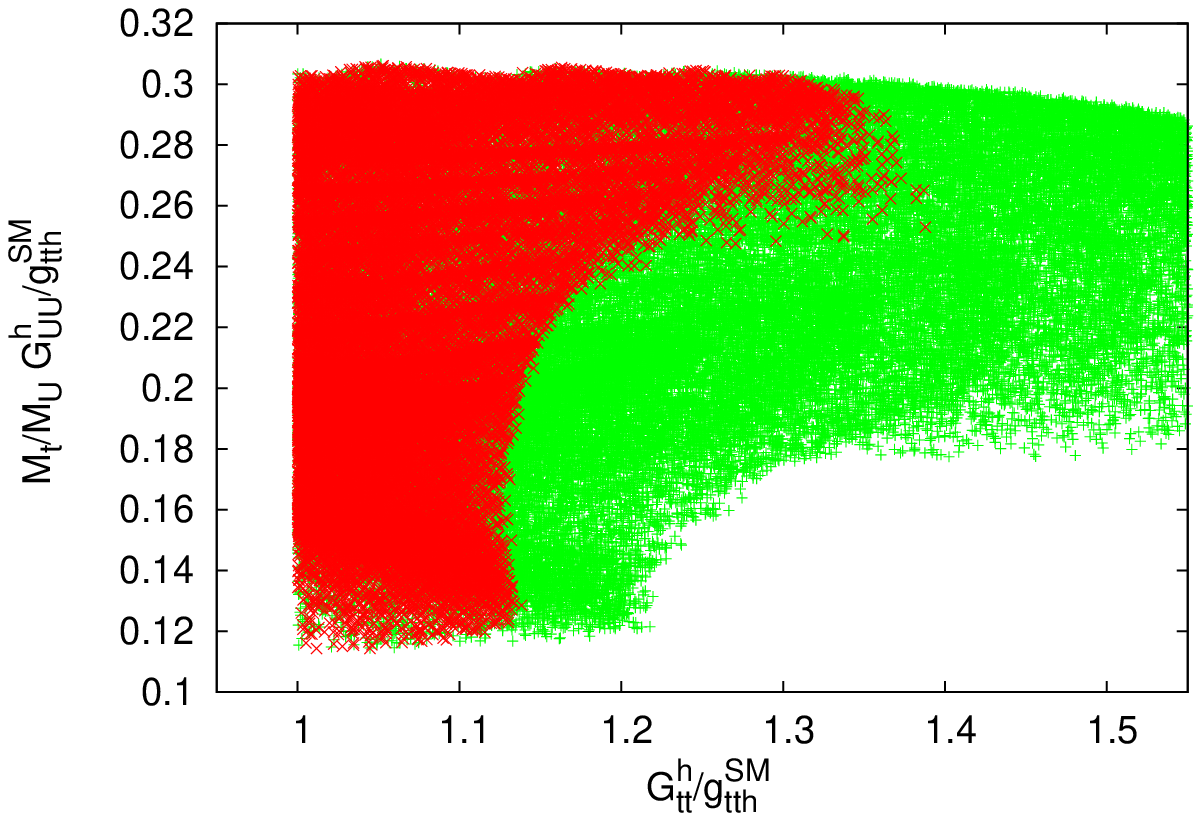}
  \end{center}
  \caption{The diagonal components of the physical yukawa couplings.
   The red points are inside of the $2\sigma$ constraints (\ref{Run1-2sigma}),
   while the green points satisfy only the conditions of 
   $G_{tt}^h/g_{tth}^{\rm SM}>1$ and $G_{TT}^h < 0$, and the $S,T$-constraints.
   \label{kT-kt} }
\end{figure}

\begin{figure}[t]
  \begin{center}
   \includegraphics[width=0.4\textwidth]{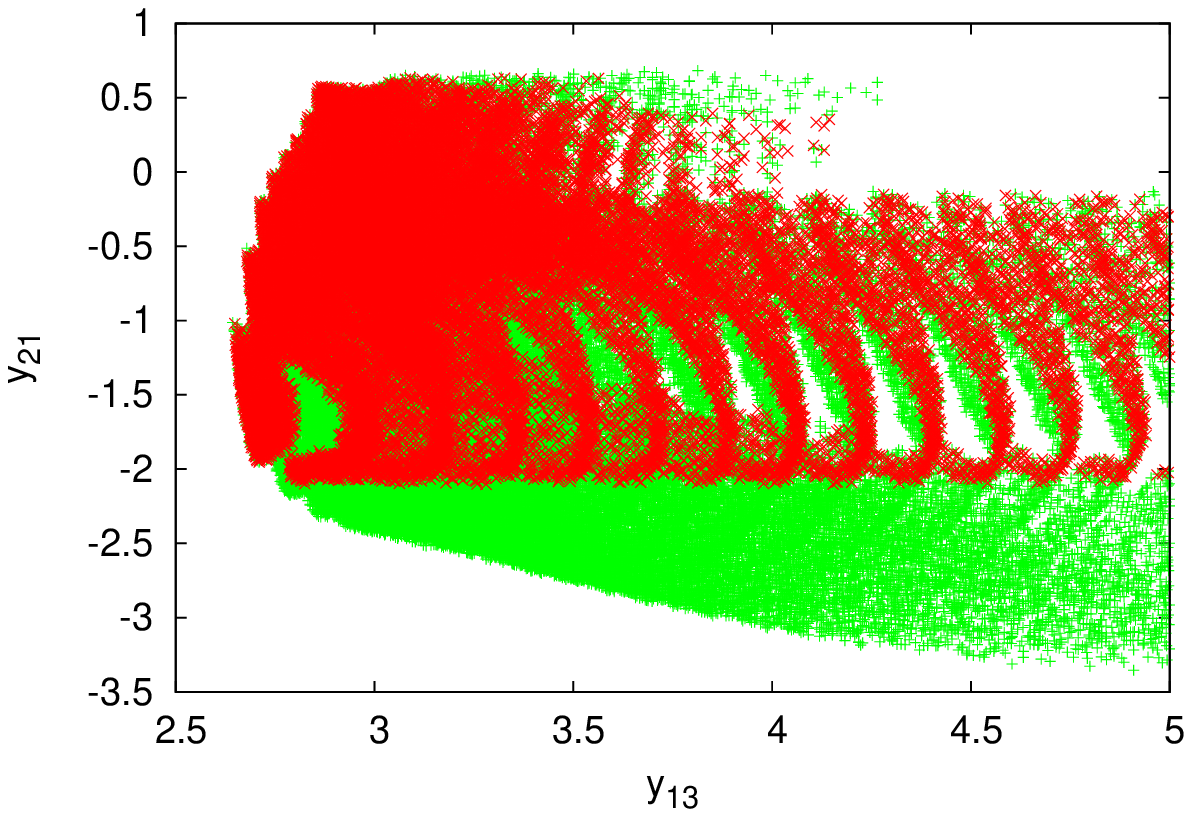}\qquad
   \includegraphics[width=0.4\textwidth]{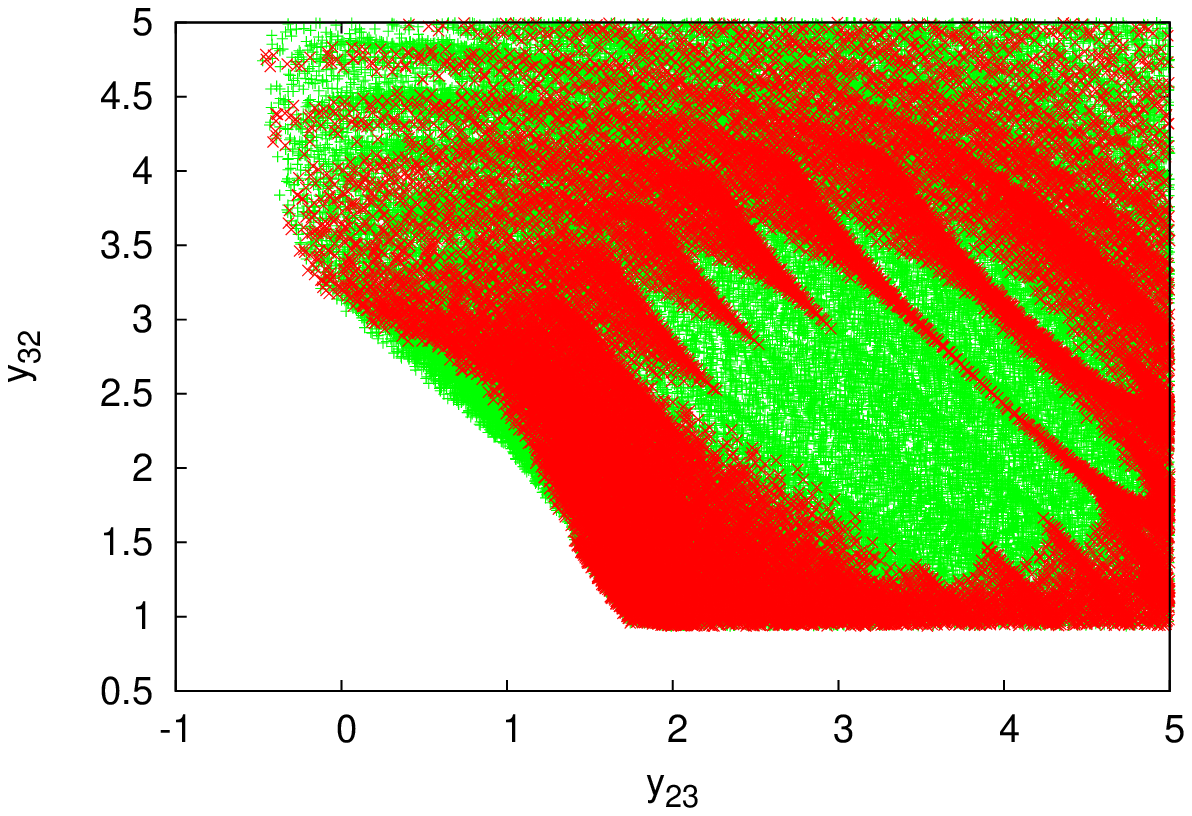}
  \end{center}
  \caption{The Lagrangian parameters of the yukawa couplings 
   in Eq.~(\ref{mass-matrix-general}).
   The red points are inside of the $2\sigma$ constraints (\ref{Run1-2sigma}),
   while the green points satisfy only the conditions of 
   $G_{tt}^h/g_{tth}^{\rm SM}>1$ and $G_{TT}^h < 0$, and the $S,T$-constraints.
   \label{fig-yij} }
\end{figure}

\begin{figure}[t]
  \begin{center}
   \includegraphics[width=0.95\textwidth]{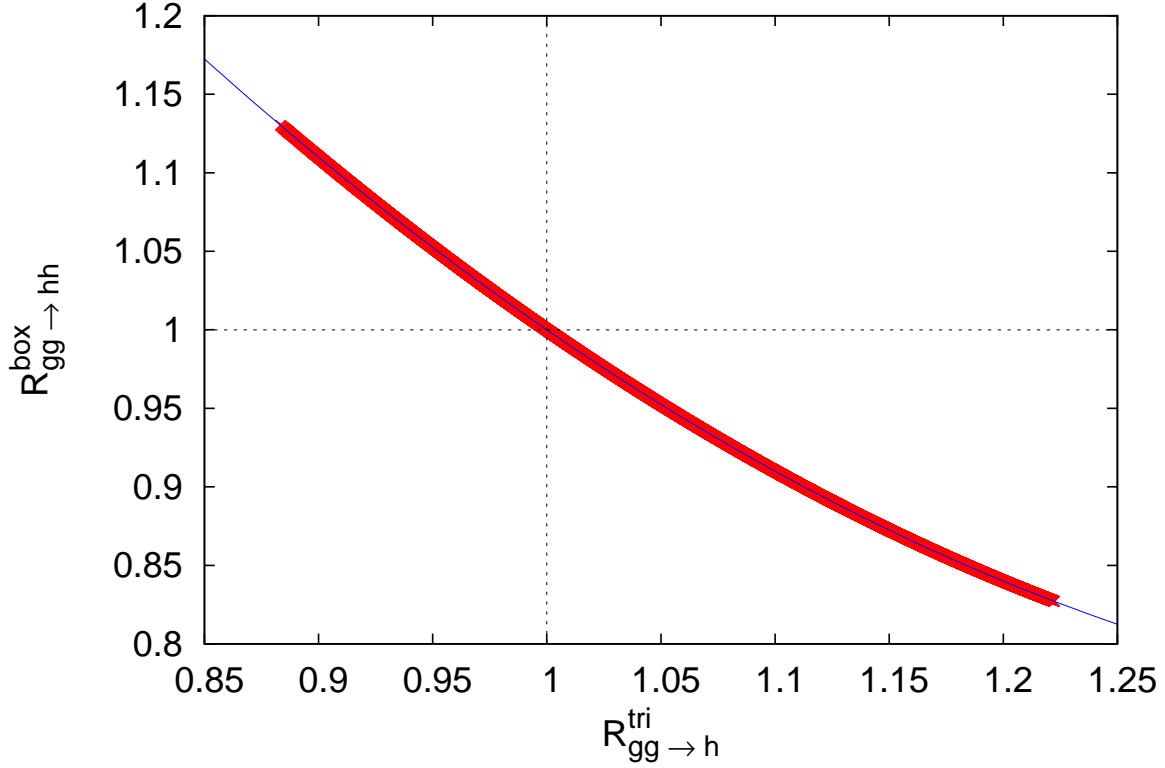}
  \end{center}
  \caption{$R_{gg \to h}^{\rm tri}$ vs $R_{gg \to hh}^{\rm box}$.
   The red points are inside of the $2\sigma$ constraints (\ref{Run1-2sigma}).
   The blue curve corresponds to the analytical relation, 
   $R_{gg \to hh}^{\rm box} = 
    \Big(R_{gg \to h}^{\rm tri}\Big)^2 - 3 \Big(R_{gg \to h}^{\rm tri}-1\Big)$, 
   shown in Eq.~(\ref{Rtri-Rbox-anal}).
   \label{Rtri-box} }
\end{figure}

\begin{figure}[t]
  \begin{center}
   \includegraphics[width=0.95\textwidth]{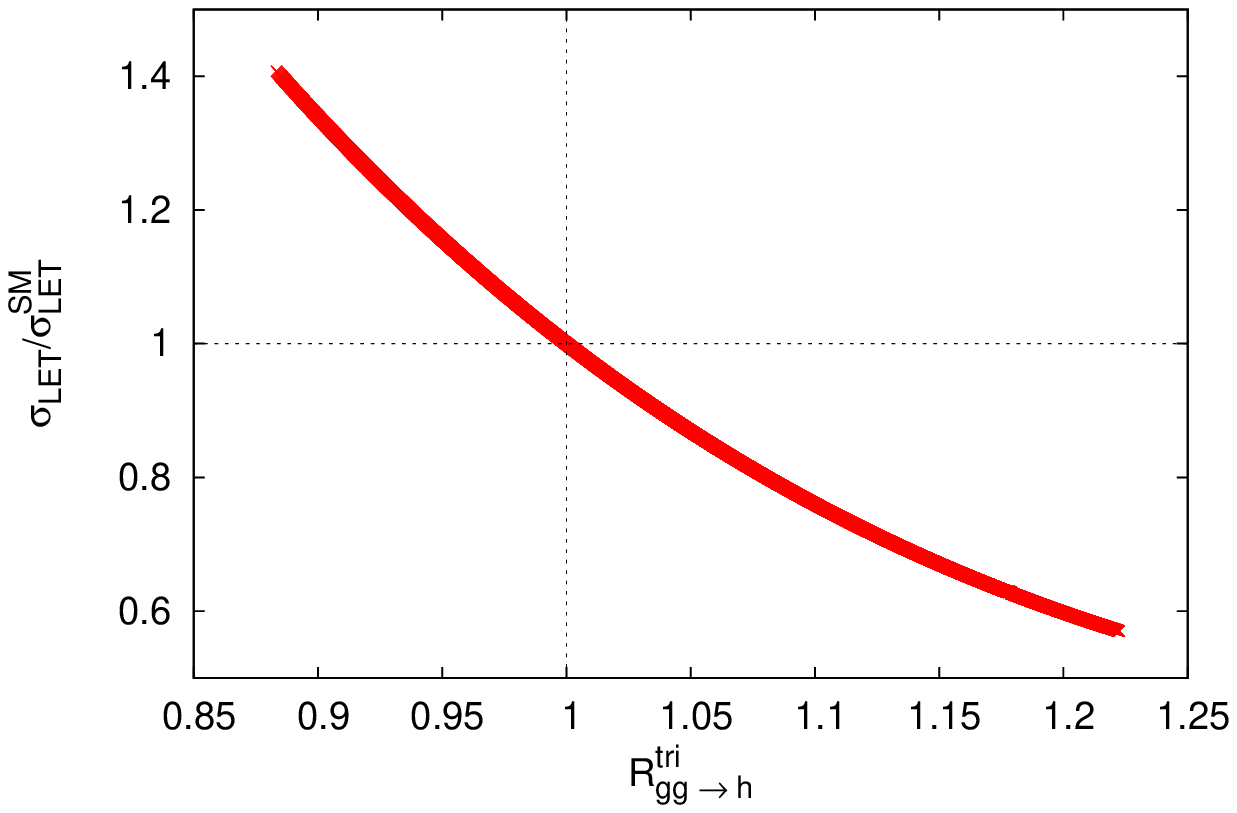}
  \end{center}
  \caption{The ratio $\sigma_{\rm LET}/\sigma_{\rm LET}^{\rm SM}$ of 
   the total $pp \to hh$ cross section at $\sqrt{s}=14$~TeV 
   normalized by the SM value under the LET approximation.
   We used the CT14 LO PDF set~\cite{Dulat:2015mca} and took 
   the renormalization and factorization scales
   equal to the invariant mass of the Higgs pair, $\mu=Q=M_{hh}=\sqrt{\hat{s}}$.
   \label{sigma_LET} }
\end{figure}

\section{Summary and discussions}
\label{Sec-summary}

We revisited the scenario with the enhanced top yukawa coupling
in the framework of the VLQ model.
We found that the scenario can be realized in the rather wide
parameter space.
Since the Lagrangian parameters of the yukawa couplings except for $y_{21}$
are positive and large, such VLQ model can be obtained from some
underlying strong dynamics.
We also calculated the ratios of the triangle and box diagrams 
to the SM values in the $gg \to hh$ process and 
found the noticeable relation.
The detailed studies will be done in future.

\appendix

\section{$S,T$-parameters in the VLQ model}

The parameter space of the VLQ models is severely restricted by
the oblique corrections~\cite{Peskin:1990zt}.
In particular, the $T$-parameter is essential.
In our model, it reads~\cite{Lavoura:1992np,Anastasiou:2009rv},
\begin{eqnarray}
  T &=& \frac{N_c}{16 \pi s_W^2 c_W^2}\bigg[\,
  \sum_{k=1}^3 \bigg(\,|L_{1k}|^2 \theta_+ (y_k,y_b)
  + (|L_{2k}|^2 + |R_{2k}|^2) \theta_+ (y_X,y_k)
  + 2 \mbox{Re}(L_{2k} R_{2k}^*) \theta_- (y_X,y_k)\,\bigg) \nonumber \\
&& \qquad
  - \frac{1}{2} \theta_+ (y_b,y_b)
  - \theta_+ (y_X,y_X) - \theta_- (y_X,y_X) \nonumber \\
&& \qquad
  - \sum_{k=1}^3 \bigg\{\,
    \frac{1}{2}\bigg((|L_{1k}|^2 - |L_{2k}|^2)^2 + |R_{2k}|^4\bigg)
    \theta_+ (y_k,y_k)
  - (|L_{1k}|^2 - |L_{2k}|^2) |R_{2k}|^2 \theta_- (y_k,y_k)\,\bigg\}
    \nonumber \\
&& \qquad
 -\sum_{i \ne j} \bigg\{\,
  \frac{1}{2} \bigg(|L_{1i}^* L_{1j} - L_{2i}^* L_{2j}|^2 + |R_{2i}^* R_{2j}|^2\bigg)
  \theta_+ (y_i,y_j) 
  \nonumber \\
&& \hspace*{2cm}
  - \mbox{Re}\bigg((L_{1i}^* L_{1j} - L_{2i}^* L_{2j})R_{2i} R_{2j}^*\bigg)
    \theta_- (y_i,y_j)\,\bigg\}\,\bigg],
\end{eqnarray}
where $N_c=3$, and 
we defined $s_W \equiv \sin \theta_W$ and $c_W \equiv \cos \theta_W$
with $\theta_W$ being the weak mixing angle, and also 
\begin{equation}
  y_i \equiv \frac{m_i^2}{m_Z^2}, \qquad
  y_b \equiv \frac{m_b^2}{m_Z^2}, \qquad
  y_X \equiv \frac{M_X^2}{m_Z^2}, 
\end{equation}
\begin{equation}
  \theta_+ (y_i,y_j) \equiv 
   y_i + y_j - \frac{2y_i y_j}{y_i-y_j}\log\frac{y_i}{y_j}
   -2(y_i \log y_i + y_j \log y_j) + \frac{y_i+y_j}{2} \Delta,
\end{equation}
and
\begin{equation}
  \theta_- (y_i,y_j) \equiv 2\sqrt{y_i y_j} \bigg(\,
  \frac{y_i+y_j}{y_i-y_j} \log \frac{y_i}{y_j} - 2 + \log (y_i y_j)
  - \frac{\Delta}{2}\,\bigg) ,
\end{equation}
with $\Delta$ being the divergent term in the dimensional regularization.
The mass eigenvalues of the up-type quarks are
$m_1 = m_t$, $m_2 = M_T$ and $m_3 = M_U$.
The rotation matrices are defined by
\begin{equation}
  \left(\begin{array}{c}t_L \\ T_L \\ U_L \end{array} \right)=
  \left(
    \begin{array}{ccc}
     L_{11} & L_{12} & L_{13} \\  L_{21} & L_{22} & L_{23} \\
     L_{31} & L_{32} & L_{33}
    \end{array}
  \right)
  \left(\begin{array}{c}t'_L \\ T'_L \\ U'_L \end{array} \right), \qquad
  \left(\begin{array}{c}t_R \\ T_R \\ U_R \end{array} \right)=
  \left(
    \begin{array}{ccc}
     R_{11} & R_{12} & R_{13} \\  R_{21} & R_{22} & R_{23} \\
     R_{31} & R_{32} & R_{33}
    \end{array}
  \right)
  \left(\begin{array}{c}t'_R \\ T'_R \\ U'_R \end{array} \right),
\end{equation}
where $(t,T,U)_{L,R}$ and $(t',t',U')_{L,R}$ are
the gauge and mass eigenstates, respectively.

We have left the divergent term $\Delta$ 
%in the dimensional regularization 
for checking of the calculations~\cite{Anastasiou:2009rv}.
By using the unitarity and
the mass relations 
\begin{equation}
  M_X = {\cal M}_{22} = (V_L \diag (m_1,m_2,m_3) V_R^\dagger)_{22}
      = m_1 L_{21} R_{21}^* + m_2 L_{22} R_{22}^* + m_3 L_{23} R_{23}^*,
\end{equation}
and also
\begin{equation}
  0 = {\cal M}_{12} = (V_L \diag (m_1,m_2,m_3) V_R^\dagger)_{12}
      = m_1 L_{11} R_{21}^* + m_2 L_{12} R_{22}^* + m_3 L_{13} R_{23}^* ,
\end{equation}
we can confirm that the divergent term $\Delta$ is exactly canceled out,
as it must be.

The deviation from the SM is given by
\begin{equation}
  \Delta T = T - T_{\rm SM},
\end{equation}
with
\begin{equation}
  T_{\rm SM} = \frac{N_c}{16\pi s_W^2 c_W^2 m_Z^2} \bigg[\, m_t^2 + m_b^2
  - 2 \frac{m_t^2 m_b^2}{m_t^2-m_b^2} \log \frac{m_t^2}{m_b^2}\,\bigg]\,.
\end{equation}
Throughout the paper, we take the $1\sigma$ constraint, 
$\Delta T=0.10 \pm 0.07$~\cite{Olive:2016xmw}. 

The $S$-parameter constraint is not so severe, compared with 
the $T$-parameter.
The expression for the $S$-parameter in our model is 
as follows~\cite{Lavoura:1992np}:
\begin{eqnarray}
  S &=& \frac{N_c}{2 \pi}\bigg[\,
  \sum_{k=1}^3 \bigg(\,|L_{1k}|^2 \psi_+ (y_k,y_b)
  + (|L_{2k}|^2 + |R_{2k}|^2) \psi_+ (y_X,y_k)
  + 2 \mbox{Re}(L_{2k} R_{2k}^*) \psi_- (y_X,y_k)\,\bigg) \nonumber \\
&& \qquad
 -\sum_{i \ne j} \bigg\{\,
  \frac{1}{2} \bigg(|L_{1i}^* L_{1j} - L_{2i}^* L_{2j}|^2 + |R_{2i}^* R_{2j}|^2\bigg)
  \chi_+ (y_i,y_j) 
  \nonumber \\
&& \hspace*{2cm}
  - \mbox{Re}\bigg((L_{1i}^* L_{1j} - L_{2i}^* L_{2j})R_{2i} R_{2j}^*\bigg)
    \chi_- (y_i,y_j)\,\bigg\}\,\bigg],
\end{eqnarray}
where we defined
\begin{equation}
  \psi_+ (y_i,y_j) \equiv \frac{1}{3} - \frac{1}{9} \log \frac{y_i}{y_j},
\end{equation}
\begin{equation}
  \psi_- (y_i,y_j) \equiv - \frac{y_i+y_j}{6\sqrt{y_i y_j}},
\end{equation}
\begin{equation}
  \chi_+ (y_i,y_j) \equiv \frac{5(y_i^2+y_j^2)-22y_iy_j}{9(y_i-y_j)^2}
   + \frac{3 y_i y_j(y_i+y_j) - y_i^3-y_j^3}{3(y_i-y_j)^3}\log\frac{y_i}{y_j},
\end{equation}
and
\begin{equation}
  \chi_- (y_i,y_j) \equiv -\sqrt{y_i y_j} \bigg(\,
  \frac{y_i+y_j}{6y_i y_j} - \frac{y_i+y_j}{(y_i-y_j)^2}+
  \frac{2y_i y_j}{(y_i-y_j)^3}\log \frac{y_i}{y_j}\,\bigg) \, .
\end{equation}
Note that $\chi_+ (y_i,y_i) = \chi_- (y_i,y_i) = 0$.

The deviation from the SM is given by
\begin{equation}
  \Delta S = S - S_{\rm SM},
\end{equation}
with
\begin{equation}
  S_{\rm SM} = \frac{N_c}{2\pi} \bigg[\,
  \frac{1}{3} - \frac{1}{9} \log \frac{m_t^2}{m_b^2}\,\bigg]\,.
\end{equation}
Throughout the paper, we take the $1\sigma$ constraint, 
$\Delta S=0.07 \pm 0.08$~\cite{Olive:2016xmw}. 

\section{Analytical expression of $R_{gg \to h}^{\rm tri}$ and 
$R_{gg \to hh}^{\rm box}$}
\label{app-b}

For the general mass matrix (\ref{mass-matrix-general})
in our model, we define the dimensionless matrices as follows:
\begin{equation}
  \widetilde{{\cal M}} \equiv \frac{1}{M_X} {\cal M} =
  \left(
    \begin{array}{ccc}
     0  & 0 & a_{13} \\ a_{21} & 1 & a_{23} \\ a_{31} & a_{32} & a_{33}
    \end{array}
   \right), \qquad
  \widetilde{\boldsymbol{Y}} \equiv
  \frac{1}{\sqrt{2}}\frac{v}{M_X}\,{\mathbf Y} = 
   \left(
    \begin{array}{ccc}
      0 & 0 & a_{13} \\ a_{21} & 0 & a_{23} \\ 0 & a_{32} & 0
    \end{array}
   \right) ,
   \label{dimless-matrix}
\end{equation}
where we scaled by $M_X = m_{22}$.
We assume $a_{13} \ne 0$ for getting $m_t \ne 0$.
We then read Eqs.~(\ref{def-Rtri}) and (\ref{def-Rbox}) as
\begin{equation}
  R_{gg \to h}^{\rm tri} = 
  \tr (\widetilde{\boldsymbol{Y}} \widetilde{{\cal M}}^{-1}), \qquad
  R_{gg \to hh}^{\rm box} =
  \tr (\widetilde{\boldsymbol{Y}} \widetilde{{\cal M}}^{-1}
       \widetilde{\boldsymbol{Y}} \widetilde{{\cal M}}^{-1}) \, .
\end{equation}
Let us determine the dimensionless mass matrix inverse 
$\widetilde{{\cal M}}^{-1}$. 
The definition of the inverse matrix,
$\widetilde{{\cal M}} \widetilde{{\cal M}}^{-1} =
 \widetilde{{\cal M}}^{-1} \widetilde{{\cal M}} = \diag(1,1,1)$, yields
the expression for $\widetilde{{\cal M}}^{-1}$ as
\begin{equation}
  \widetilde{{\cal M}}^{-1} = 
   \left(
    \begin{array}{ccc}
     b_{11}  & b_{12} & b_{13} \\ b_{21} & b_{22} & b_{23} \\
     \frac{1}{a_{13}} & 0 & 0
    \end{array}
   \right),
\end{equation}
and the matrix elements $b_{ij}$ satisfy
\begin{equation}
  \left(
    \begin{array}{cc}
     a_{21} & 1 \\ a_{31} & a_{32}
    \end{array}
   \right)
   \left(
     \begin{array}{cc}
       b_{12} & b_{13} \\ 
       b_{22} & b_{23} 
     \end{array}
    \right) = 
   \left(
     \begin{array}{cc}
       1 & 0 \\ 0 & 1
     \end{array}
    \right), \qquad 
  \left(
    \begin{array}{cc}
     a_{21} & 1 \\ a_{31} & a_{32}
    \end{array}
   \right)
   \left(
     \begin{array}{c}
       b_{11}  \\ b_{21}
     \end{array}
    \right) = -\frac{1}{a_{13}}
   \left(
     \begin{array}{c}
      a_{23}  \\ a_{33}
     \end{array}
   \right) \, .
\end{equation}
Therefore we analytically obtain $b_{ij}$, 
\begin{equation}
   \left(
     \begin{array}{cc}
       b_{12} & b_{13} \\ 
       b_{22} & b_{23} 
     \end{array}
    \right) = \frac{1}{a_{21}a_{32}-a_{31}}
  \left(
    \begin{array}{cc}
     a_{32} & -1 \\ -a_{31} & a_{21}
    \end{array}
   \right), 
\end{equation}
and
\begin{equation}
   \left(
     \begin{array}{c}
       b_{11}  \\ b_{21}
     \end{array}
    \right)
  % = -\frac{1}{a_{13}(a_{21}a_{32}-a_{31})}
  % \left(
  %   \begin{array}{cc}
  %    a_{32} & -1 \\ -a_{31} & a_{21}
  %   \end{array}
  %  \right)
  %  \left(
  %    \begin{array}{c}
  %      a_{23}  \\ a_{33}
  %    \end{array}
  %   \right)
   = - \frac{1}{a_{13}(a_{21}a_{32}-a_{31})}
    \left(
     \begin{array}{c}
       a_{32} a_{23} - a_{33} \\ -a_{31} a_{23} + a_{21} a_{33}
     \end{array}
    \right) \, . 
\end{equation}
By using $a_{21} b_{12} = a_{32} b_{23} = 1-b_{22}$,
we find
\begin{equation}
  R_{gg \to h}^{\rm tri} = 3 - 2 b_{22} ,
\end{equation}
and
\begin{equation}
  R_{gg \to hh}^{\rm box} = 1 + 2 (1-b_{22})^2 - 2 b_{22} (1-b_{22})
  = 3-6b_{22} + 4b_{22}^2 \, .
\end{equation}
Eliminating $b_{22}$ from the above equations, we obtain 
Eq.~(\ref{Rtri-Rbox-anal}),
\begin{equation}
  R_{gg \to hh}^{\rm box} = 
  \Big(R_{gg \to h}^{\rm tri}\Big)^2 - 3 \Big(R_{gg \to h}^{\rm tri}-1\Big) \, .
\end{equation}
When $a_{21}=0$ as in Sec.~\ref{crude-app},
we immediately find $b_{22}=1$ and thereby obtain
$R_{gg \to h}^{\rm tri} = R_{gg \to hh}^{\rm box} = 1$.

\acknowledgments

The author thank to A. Deandrea and G. Cacciapaglia for useful comments.
Numerical computation in this work was carried out at 
the Yukawa Institute Computer Facility.
This work is supported by JSPS Grant-in-Aid for Scientific Research 
No.~17K05423 and partially by the France-Japan Particle Physics Lab 
(TYL/FJPPL).

\end{document}